\documentstyle[12pt]{article}
\def\beq{\begin{equation}}
\def\eeq{\end{equation}}
\def\bea{\begin{eqnarray}}
\def\eea{\end{eqnarray}}
\def\bq{\begin{quote}}
\def\eq{\end{quote}}
\def\PL{{\it Phys. Lett.} }
\def\PRL{{\it Phys. Rev. Lett.} }
\def\NP{{\it Nucl. Phys.} }
\def\PR{{\it Phys. Rev.} }

\def\gappeq{\mathrel{\rlap {\raise.5ex\hbox{$>$}}
{\lower.5ex\hbox{$\sim$}}}}

\def\lappeq{\mathrel{\rlap{\raise.5ex\hbox{$<$}}
{\lower.5ex\hbox{$\sim$}}}}
\begin{document}
\topmargin -0.5cm
\oddsidemargin -0.8cm
\evensidemargin -0.8cm
\pagestyle{empty}

\vspace*{2.0cm}
\begin{center}
{\bf SUPERGRAVITY AND THE QUEST FOR A UNIFIED THEORY}\\
\vspace*{2cm}
Sergio Ferrara\\
\vspace*{0.5cm}
Theoretical Physics Division, \\
CERN, Geneva, Switzerland\\
\vspace {2.0cm}

{\bf Abstract}
\end{center}
A recollection of some theoretical developments that preceded and
followed the first formulation of supergravity theory is presented.
Special emphasis is placed on the impact of supergravity on the search
for a
unified theory of fundamental interactions.
\vspace{3.0cm}
\begin{center}
 {\it Dirac Lecture delivered at \\
the International Centre for Theoretical Physics, Trieste,\\
 19 April 1994}
\end{center}
\pagestyle{empty}
\clearpage\mbox{}\clearpage
\setcounter{page}{1}
\pagestyle{plain}
\vfill\eject

It is a great honour and pleasure to be invited to give this Dirac
Lecture on the
occasion of the 1994 Spring School on String, Gauge Theory and Quantum
Gravity.

In fact, this School is a continuation of a very successful series
initiated by
Prof.~A.~Salam in 1981.  Together with J.G.~Taylor and
P.~van~Nieuwenhuizen I
had the privilege of organizing the first two in the spring of 1981 and
the fall of 1982 \cite{aaa}.

At that time, supergravity was in the mainstream of research, namely
\begin{itemize}
\item[1)] Quantum properties of extended supergravities and their
geometric
structure,
\item[2)] Kaluza--Klein supergravity,
\item[3)] Models for particle physics phenomenology.
\end{itemize}
These topics were widely covered during the first two schools and
workshops.

Before going on to discuss supergravity and its subsequent development,
let me
briefly touch upon the steps taken in the two preceding years, when
supersymmetry
in four dimensions was introduced.

Although the latter, with its algebraic structure, was first mentioned
in 1971 by
Gol'fand and Likhtman \cite{bb} and in early 1973 by Volkov and Akulov
\cite{cc} (to
explain the masslessness of the neutrino as a Goldstone fermion), it was
really
brought to the attention of theoretical particle physicists in the
second half of
1973, by Wess and Zumino \cite{dd};  they had been inspired by a similar
structure,
found by Gervais and Sakita (1971) \cite{ee}, already present in two
dimensions, in
the dual-spinor models constructed in 1971 by Neveu and Schwarz
\cite{ff} and by
Ramond \cite{ggg}.  The relevance of supersymmetry for quantum field
theory,
especially in view of its remarkable ultraviolet properties and its
marriage with
Yang--Mills gauge invariance, was soon established in early 1974.

It is nevertheless curious that it was only, at the time, rather
isolated groups that
delved into the subject, mainly in Europe:  at CERN, the ICTP (Trieste),
Karlsruhe,
the ENS-Paris, Imperial College-London, Turin Univ., and essentially two
in the United
States:  Caltech and Stony Brook. The same applies to supergravity and
its
ramifications in the early years, after its foundation in 1976.

Soon after the very first paper of Wess and Zumino \cite{hh}, a
remarkable sequence of events occurred during 1974:
\begin{itemize}
\item The superspace formulation of supersymmetric field theories
(Salam, Strathdee \cite{jj};  Wess, Zumino, Ferrara \cite{kk}).
\item The discovery of non-renormalization theorems (Wess, Zumino
\cite{lll};
Iliopoulos, Zumino \cite{mm};  Ferrara, Iliopoulos, Zumino \cite{nn}).
\item The construction of supersymmetric Yang--Mills theories (Wess,
Zumino \cite{oo}
for the Abelian case;  Ferrara, Zumino \cite{pp}, and Salam, Strathdee
\cite {qq}, for
the non-Abelian case). \item The first construction of a renormalizable
field theory
model with spontaneously broken supersymmetry (Fayet, Iliopoulos
\cite{rr}).
 \item The construction of
a multiplet of currents, including the supercurrent and the stress
energy tensor (Ferrara, Zumino \cite{ss}), which act as a source for the
supergravity
gauge fields and had an impact also later, in the classification of
anomalies and in
the covariant construction of superstring Lagrangians.
\end{itemize}

In the same year, quite independently of supersymmetry, Scherk and
Schwarz \cite{tt}
proposed string theories as fundamental theories for quantum gravity and
other gauge forces rather than for hadrons, turning the Regge slope from
$\alpha'
\sim$ GeV$^{-2}$ to $\alpha' \sim 10^{-34}$~GeV$^{-2}$, the evidence
being that
any such theory contained a massless spin 2 state with interactions for
small
momenta as predicted by Einsteins' theory of general relativity.

In the following year, many models with spontaneously broken
supersymmetry and
gauge symmetry were constructed, mainly by Fayet \cite{uu} and
O'Raifeartaigh
\cite{vv}, and $N = 2$ gauge theories coupled to matter, which were
formulated by
Fayet \cite{ww}.

This was a prelude to two importants events, which took place just
before  and soon
after
 the proposal of supergravity: the discovery of extended superconformal
algebras (Ademollo et al., Nov. 1975 \cite{yy}) and the finding of
evidence for
space-time supersymmetry in superstring theory (Gliozzi, Olive, Scherk,
GOS, for
short, Sept. 1976 and Jan. 1977 \cite{zz}).  In retrospect, these
episodes had a great
impact on the subsequent development of string theories in the mid 80's:

The Ademollo et al. paper, just a few months before
supergravity was formulated, was inspired by the fact that it was
possible, in
higher dimensions ($D = 4$), by undoing the superspace coordinate
$\theta_i$ with
a counting index $(i = 1,...,N)$, to construct extended supersymmetries;
indeed,
a remarkable theory with $N$ = 2 ($D$ = 4) supersymmetry then had just
been
discovered by Fayet (Sep. 1985 \cite{ww}).  In $D$ = 4, extended
superconformal
algebras were accompanied by U($N$) gauge algebras [SU(4) for $N$ = 4].
In $D$ = 2,
superconformal algebras are infinite-dimensional and $N$ = 2 and $N$ = 4
turned out
to be accompanied by U(1) and SU(2) Kac--Moody gauge algebras.   These
algebras, at
the time thought of as gauge-fixing of superdiffeomorphisms, were
introduced to study
new string theories with different critical dimension \cite{aai}.  In
retrospect, this
construction had a major impact on the classification of ``internal"
superconformal
field theories, especially $N$ = 2, as the quantum version of
Calabi--Yau manifolds,
and on its relation \cite{bbi} with the existence of space-time
supersymmetry
in $D < D_{\rm crit}$.

Meanwhile, in the spring of 1976 \cite{cci}, supergravity was
formulated by Freedman, van Nieuwenhuizen and the author, working at the
Ecole
Normale and at Stony Brook.  Soon after, a simplified version
(first-order
formulation) was presented by Deser and Zumino \cite{ddi}.
\vfill\eject
While in the second formalism, the spin-3/2 four-fermion interaction
has the meaning of a contact term (similar to seagull terms in scalar
electrodynamics
or non-Abelian gauge theories) required by fermionic gauge invariance,
in the
first-order formalism it has the meaning of a torsion contribution to
the spin
connection from ``spin-3/2 matter".  This discovery also implied that
any
supersymmetric system coupled to gravity should manifest local
supersymmetry.

This observation eventually led some physicists to go deeper in string
theory in
order to explore whether the ``dual spinor model" could accommodate
target-space supersymmetry.  The GOS paper (Sept. 1976 and Jan. 1977
\cite{zz}) gave
dramatic evidence for space-time supersymmetry in the superstring theory
(called at
that time the dual-spinor model) \cite{ddia} by cutting out the $G$-odd
parity states
in the N-S sector and comparing its bosonic spectrum with the fermion
spectrum of the
Ramond sector.  In the proof, they used an identity that had been proved
by Jacobi in
1829 {\it(Aequatio identica satis abstrusa)}!  This paper came out
following some
sequential developments in supergravity, just after its first
construction in the
spring of 1976, namely the first matter-coupling to Maxwell theory
(Ferrara, Scherk,
van Nieuwenhuizen, Aug. 1976 \cite{eei}), to Yang--Mills theory and
chiral
multiplets \cite{ffi} and the first formulation of extended supergravity
[$N$ = 2]
(Ferrara, van Nieuwenhuizen, Sept. 1976 \cite{ggi}).  It is interesting
to note
that two of the GOS authors (G and S) also took part in some of the
above-mentioned
supergravity papers.

The hypothesis of GOS (later proved in great detail by Green and Schwarz
\cite{hhi})
also implied the existence of an $N$ = 4 Yang--Mills theory, eventually
coupled to
an $N$ = 4 extended supergravity.  This was implied by a dimensional
reduction
of the 10$D$ spectrum.  The full $N$ = 4 supergravity contained in this
reduction
was found a year later (Cremmer, Scherk, Ferrara, Dec. 1977 \cite{jji})
and it was
shown to contain  an SU(4) $\times$ SU(1,1) symmetry.  Meanwhile, three
other
important developments were announced at the end of 1976.
The construction of $N$
= 3 supergravity (Freedman; Ferrara, Scherk, Zumino,  Nov. 1976
\cite{kki}) and
the discovery of (Abelian and non-Abelian) duality symmetries,
generalizing the
electromagnetic duality  F $\rightarrow \tilde {\rm F}$ in $N$ = 2
[U(1)] and $N$ = 3
[U(3)] supergravity (Dec. 1976 \cite{lli}).  This duality generalizes to
SO(6) $\times$
SU(1,1) in pure $N$ = 4 supergravity and to SO(6,$n) \times$ SU(1,1) in
$N$ = 4
supergravity coupled to $n$ matter (Yang--Mills) multiplets.

In retrospect, these symmetries play a crucial role in compactified
superstrings,
where the manifold
$$
\frac{{\rm SO(6},N)}
{{\rm SO(6)}\times{\rm SO}(N)} \times
\frac{{\rm SU(1,1)}}{{\rm U(1)}}
$$
(modded out by some discrete symmetries) describes the moduli space of
toroidally
compactified 10$D$ strings, according to the analysis of Narain,
Sarmadi,
Witten \cite{mmi}).

In September 1976, also the covariant world-sheet formulation of
the spinning string was presented in two papers \cite{nni} by Brink, Di
Vecchia and
Howe and by Deser and Zumino.  In retrospect this can be considered as a
crucial
ingredient for the Polyakov formulation \cite{ooi} of spinning strings
with arbitrary
world-sheet topology. In this respect, ($p$ + 1) supergravity is
necessary for the
consistent formulation of any $p$-dimensional extended object coupled to
fermions.

In the subsequent years all higher extended 4$D$ supergravities with $N$
= 5, 6 and 8
were constructed.

The maximally extended supergravity ($N$ = 8) was found by Cremmer and
Julia
\cite{ppi}, by dimensional reduction of 11$D$ supergravity previously
obtained by the
same authors with Scherk (1978 \cite{qqi}), and its gauged version,
accompanied with an
SO(8) Yang--Mills symmetry, by de Wit and Nicolai (1982 \cite{rri}).
Gell-Mann had
earlier observed that SO(8) cannot accommodate the observable gauge
symmetry SU(3)
$\times$ SU(2) $\times$ U(1) of electroweak and strong interactions.
However, it was
later observed by Ellis, Gaillard, Maiani and Zumino (1982 \cite{ssi})
that a hidden
local SU(8) symmetry (found by Cremmer and Julia) could be identified as
a viable Grand
Unified Theory (GUT) for non-gravitational interactions.  The basic
assumption was
that the degrees of freedom of the SU(8) gauge bosons could be generated
at the
quantum level, as it was known to occur in certain 2$D$
$\sigma$-models, following the analysis of Di~Vecchia, D'Adda and
L\"uscher
\cite{tti}.  However, contrary to 2$D~\sigma$-models, which are
renormalizable and
therefore consistent quantum field theories, it turned out later that
$N$ = 8
supergravity in $D$ = 4, which is also a kind of generalized
$\sigma$-model, is
unlikely to enjoy a similar property.  This is one of the reasons why
this attempt was
abandoned.  Another reason was closely related to the forthcoming string
revolution,
when Green and Schwarz (GS) (1984 \cite{uui}) proved that $D$ = 10, $N$
= 1
supergravity, coupled to supersymmetric Yang--Mills matter, could be
embedded in a
consistent superstring theory for a particular choice of gauge groups
(SO(32) and E$_8
\times$E$_8$).

The GOS and GS papers gave strong evidence that superstrings consistent
with space-time supersymmetry containing supergravity + matter (rather
than pure
higher extended supergravity), in the massless sector, were a possible
candidate for a
theory of quantum gravity, encompassing the other gauge interactions and
free from
unphysical ultraviolet divergences.  On the contrary, in the context of
point-field
theories, these systems, even if the symmetries dictated in an almost
unique way all
the couplings, were found to be non-renormalizable, already at one loop,
when standard
perturbative techniques were applied to them (Grisaru, van
Nieuwenhuizen, Vermaseren,
1976 \cite{vvi}).  Indeed it was later shown that this was also the case
for pure
supergravities at and beyond three loops.  [These theories had, however,
the
remarkable property of being one- and two-loop finite (Grisaru, van
Nieuwenhuizen,
Vermaseren \cite{vvi}; Grisaru \cite{wwi}; Tomboulis \cite{wwi}).]
Pioneering work, in
the late 70's, was also the analysis of spontaneous supersymmetry
breaking in global
and local supersymmetry.  In rigid supersymmetry, Fayet \cite{yyi}
opened the way
to the construction of the minimal supersymmetric extension of the
Standard Model
(MSSM), which in particular demanded two Higgs doublets.  However, the
gauge and
supersymmetry breaking introduced by him required more degrees of
freedom than
the MSSM.

When supersymmetry is gauged, i.e. in supergravity, the supersymmetric
version of
the Higgs mechanism appears (super-Higgs), i.e. the goldstino is eaten
up by the
spin-3/2 gravitino (the gauge fermion of supergravity, the partner of
the
gravitons), which then becomes massive.

The possibility of having spontaneously broken supergravity with
vanishing
cosmological constant was shown by Deser and Zumino  (Apr. 1977
\cite{zzi}) and
proved in detail by Cremmer\break\hfill
\newpage
\noindent
et al. (Aug. 1978 \cite{aaii}), by studying the most
general matter coupling to $N$ = 1 supergravity for a chiral multiplet,
whose
superpotential triggers a non-vanishing gravitino mass.  The Higgs
effect for
Goldstone fermions had also been considered earlier by Volkov and Soroka
\cite{bbii}.

Another important result at that time, found by Zumino (Aug. 1979
\cite{ccii}), was
the fact that the most general supergravity couplings of chiral
multiplets (with
two-derivative action) were described by K\"ahlerian $\sigma$-models.

Again, in retrospect, this K\"ahlerian structure and the generalization
thereof
have played a role in superstring theory from both the world-sheet and
target-space
points of view.

Although in the 70's the work done in supersymmetric models for particle
physics
(using renormalizable Lagrangians with spontaneously broken
supersymmetry) and that
towards a deeper understanding of the structure of supergravity theories
(off-shell
formulations, matter couplings, etc.) went in parallel, with small
intersections,
they came closer and became eventually deeply connected after two major
developments were made in the early 80's.

The first was the call made upon supersymmetry breaking near the
electroweak scale, to
solve the so-called hierarchy problem of gauge theories with fundamental
Higgs scalars
(Gildener, Weinberg; Veltman; Witten; Maiani) \cite{ddii},\cite{eeii}.

This development and general properties of criteria for supersymmetry
breaking,
contained in two pivotal papers by Witten (Apr. 1981, Jan. 1982
\cite{eeii}), opened up
the field of supersymmetry and supergravity as main stream research in
the United
States and in the rest of the world.

The hierarchy problem is connected to the unnaturalness of the hierarchy
$E_{\rm
F}/E_X$ ($E_{\rm F}$ being the Fermi scale) in any renormalizable theory
with
fundamental scalars, whose v.e.v. triggers the electroweak gauge
symmetry breaking at a
scale $E_{\rm F}$ much lower than any other (cut-off) scale $E_X$.

This is due to the quadratic dependence on the cut-off $\Lambda$ of the
effective
potential, which, at one loop, manifests itself in a term
$$
\sum_{J_i} (-)^{2J_i}(2J_i+1) {\cal M}^2_{J_i}(\phi)\Lambda^2~.
$$
where $M^2_{J_i}(\phi)$ are the (scalar) field-dependent masses of
particle species
with spin $J_i$. In an arbitrary supersymmetric renormalizable field
theory with no
traceful Abelian gauge group factor, the expression multiplying
$\Lambda^2$
identically vanishes (owing to the special relation between boson and
fermion
couplings), as was shown by Girardello, Palumbo and the author (Apr.
1979
\cite{ffii}). This is also true for matter-coupled $N$ = 1 supergravity
with a
single chiral multiplet on a flat K\"ahler manifold (1978 \cite{aaii})
and in
spontaneously broken extended supergravity via the Scherk--Schwarz
mechanism (1979
\cite{ggii}).

However, a closer look at boson--fermion mass matrices revealed that
this property
made models previously considered by Fayet more problematic, since they
tended
either to give an unrealistic spectrum with some scalar superpartners of
quarks and
leptons lighter than their fermionic counterparts, or to need a traceful
additional U(1) gauge interaction, plagued with triangular anomalies.
Cancelling
these anomalies usually needed extra fields, which eventually allowed
vacua with
broken colour or charge symmetry.

However, when the most general coupling of $N$ = 1 supergravity to an
arbitrary
matter system, with arbitrary gauge interactions, became available
(Cremmer,
Ferrara, Girardello, Van Proeyen, 1982 \cite{hhii}), it was realized
that, provided
$m_{3/2} \ll M_{\rm Pl}$ and possibly $\simeq O$(TeV), mass terms for
any observable
scalar $O(m_{3/2})$ were easily generated, thus resolving the
partner--spartner
splitting problem, which generally occurred in spontaneously broken
rigid theories.

There is an alternative way of phrasing this:  in the Fayet-type models,
the
goldstino has coupling to the observable sector $O$(1) and the gravitino
mass
is very tiny, $m_{3/2} \sim 10^{-13}$~GeV, while in supergravity models
with $m_{3/2}
\gappeq O(m_{\rm Z})$ the goldstino coupling is highly suppressed
$[O(m_{3/2}/M_{\rm
Pl})]$, which implies that the gravitino only carries gravitational
interactions
(Fayet \cite{jjii}).

In the limit in which $m_{3/2}$ is kept fixed and couplings $O(1/M_{\rm
Pl})$ are
neglected, spontaneously broken supergravity models behave as globally
supersymmetric models with softly broken terms, i.e. terms with
dimension $\leq
3$, which do not induce quadratic divergences in the low-energy
effective theory.

These terms had been classified in 1981 by Girardello and Grisaru
\cite{kkii}.  A
generalization of non-renormalization theorems for superpotential terms
in a
generic theory were also found using superspace techniques, by Grisaru,
Siegel,
Ro$\check {\rm c}$ek (June 1979 \cite{llii}).

 Softly broken terms and
renormalization theorems were used to construct viable supersymmetric
GUTs,
including the MSSM as their low-energy effective theory, with no
hierarchy problem
(the first of these was constructed by Georgi and Dimopoulos in the
summer of 1981
\cite{mmii}).  Soon after, realistic electroweak and GUT models, with
spontaneously broken supersymmetry triggered by the supergravity
couplings at the
tree level, were constructed (Barbieri, Ferrara, Savoy, 1982
\cite{nnii};
Chamseddine, Nath, Arnowitt, 1982 \cite{ooii}; Hall, Lykken, Weinberg
1983
\cite{ppii}).  A general feature of these models is that the messengers
of
supersymmetry breaking to the observable sector (encompassing
electroweak and
strong interactions) are a set of neutral chiral multiplets (called the
hidden
sector), which have only gravitational interactions and decouple from
the
low-energy theory;  in the latter, the only trace of them is to produce
the
soft-breaking terms, then having the effect of modifying the supertrace
formula of
global sypersymmetry with an additional (field-independent) constant
(with no
physical consequences on the theory decoupled from gravity).

Nowadays, in the MSSM, the electroweak symmetry is broken through
radiative
corrections, through a Coleman--Weinberg mechanism, while supersymmetry
is broken
at the tree level through the soft-breaking terms.

Considering the initial condition for the couplings as given at the
Planck scale
and evolving them through the renormalization group equations
(Ib\`a$\tilde {\rm
n}$ez, Ross in 1981 \cite{qqii}, Alvarez-Gaum\'e, Polchinski, Wise in
1982
\cite{rrii}), in a region of the parameter space, the electroweak
symmetry is indeed
found to be spontaneously broken with a Higgs mass of the same order of
magnitude as
the gravitino mass.    There is a particular subclass of spontaneoulsy
broken
supergravity models, called no-scale supergravities (Cremmer, Ferrara,
Kounnas and Nanopoulos, 1983 \cite{ssii}; Ellis et al. \cite{ttii}),
where the
supergravity-breaking scale $m_{3/2}$ is arbitrary at the tree level
(owing to
flat directions in the superpotential).  In these models, radiative
corrections
may generate the hierarchy $m_{3/2} = M_{\rm Pl}~~{\rm e}^{-c/g^2}$,
then
explaining how a small scale can be created in a theory whose only
original
dimensionful scale is $M_{\rm Pl}$.

It was later shown by Witten \cite{uuii} that many $4D$ superstring
models have, in the
point-field limit, a no-scale structure;  therefore, after supersymmetry
breaking,
they may give rise to a dynamical hierarchy.

 Nowadays almost every particle physicist
knows what $\tan \beta,A, B$ represent in the general parametrization of
the
soft-breaking terms of the MSSM.

The second breakthrough was on physics at the Planck scale (Green,
Schwarz, Sept. 1984
\cite{uui}), namely the discovery of anomaly-free 10$D$ supergravity
coupled to
Yang--Mills matter or consistent superstring theories, for specific
gauge group
choices (in open and heterotic strings) (Gross, Harvey, Martinec, Rohm,
Nov. 1984
\cite{vvii}).  Heterotic string theories, after suitable
compactification of six extra
dimensions (on some  compact manifolds with special properties), led to
$N = 1$
effective supergravity theories, with a spectrum of charged chiral
multiplets (chiral
with respect to the surviving gauge group $G' \supset$ SU(3) $\times$
SU(2) $\times$
U(1) (after compactification) and accommodating families of quarks and
leptons with the
SU(3) $\times$ SU(2) $\times$ U(1) assignment of the Standard Model.

The use of $10D$ Yang--Mills fields, prior to compactification, is
crucial to overcome
previous difficulties encountered in Kaluza--Klein supergravities
(Freund, Rubin,
1980 \cite{wwii};  Witten, 1981 \cite{yyii};  Duff, Nilsson, Pope, 1986
\cite{zzii}),
where attempts were made at obtaining the SU(3) $\times$ SU(2) $\times$
U(1) gauge
symmetries from the isometries of the internal manifold.  In fact, even
if in some
cases the desired gauge group could be obtained (Witten, 1981
\cite{yyii};  Castellani,
D'Auria, Fr\'e, 1984 \cite{aaiii}), these failed because the resulting
fermion spectrum
was not chiral with respect to the electroweak gauge symmetry.

 In models where supersymmetry breaking occurs via a non-trivial dilaton
superpotential, the neutral fields coming from the internal components
of the metric
tensor ~$G_{IJ}$ are natural candidates for flat directions, at least in
the limit of
manifolds that are ``large" with respect to the string scale.

In particular, in $4D$ heterotic superstring theories, compactified on
Calabi--Yau
manifolds (Candelas, Horowitz, Strominger, Witten, Jan. 1985
\cite{bbiii}), or their
``orbifold limit"\cite{bbiiia}, a natural identification of the hidden
sector occurs
with a set of ``moduli fields", which parametrize the deformations of
the K\"ahler
class and complex structure of the manifold (generalization of radial
deformations of
simple tori) \cite{cciii}.  A popular scenario for a non-perturbative
dilaton
superpotential is the gaugino-condensation mechanism (Ferrara,
Girardello, Nilles
\cite{ddiii}) in the hidden sector gauge group (Derendinger,
Ib\'a$\tilde{\rm n}$ez,
Nilles;  Dine, Rohm, Seiberg, Witten \cite{eeiii}).  The fact that some
moduli remain
large (with respect to the string scale) could result in a sliding
gravitino mass,
which could then be stabilized through radiative corrections in the
observable sector
with a dynamical suppression with respect to $M_{\rm Pl}$.

In recent years a suggestion has been made (Duff \cite{ffiii};
Strominger
\cite{ggiii}) that strings may, in the strong coupling regime, have a
simpler
formulation in terms of a dual theory (five-brane) in the weak coupling.
These
theories, in $D = 10$, have the same field theory limit, namely, $D =
10$ supergravity
(which is unique because of supersymmetry), but electrically charged
massive string
states correspond to ``magnetically" charged five-brane states (and vice
versa) with a
similar relation as it occurs between electric and magnetic charge in
Dirac monopole
quantization \cite{hhiii}.

This is an explicit manifestation of the general fact that a $(p + 1)$
form gauge
field is, in $D$ dimensions, naturally coupled to a $p$-dimensional
extended
object, and that its ``dual" potential (which is a $D-p-3$ form) is
naturally coupled
to a $D-p-4$ extended object.  From topological arguments, similar to
Dirac's, the
product of the two couplings must be quantized.

In toroidal compactifications, it has indeed been shown (Sen and Schwarz
\cite{jjiii})
that the spectrum of both electrically and magnetically charged states
(the latter
obtained from the low-energy effective field theory, i.e. an $N = 4$
supergravity
coupled to Yang--Mills) satisfies an SL(2,Z) duality for the dilaton
chiral
multiplet $S = (1/g^2) + i\theta~(g^2$ is the field-dependent gauge
coupling and
$\theta$ is the field-dependent ``$\theta$-angle").  This may
therefore suggest a ``modular-invariant" dilaton potential \cite{kkiii}.
This symmetry is similar to the ``moduli duality", which occurs in
weakly coupled
strings as a consequence of the world-sheet non-trivial topology
\cite{lliii}.

This approach is worth exploring, even if it is difficult to imagine
that a
microscopic, consistent quantum theory describing five-brane propagation
could
possibly exist.

Finally, let me conclude by making some remarks about the possible
indirect
experimental signals, indicating that supersymmetry may be just around
the corner.
With an optimistic attitude, these are
\begin{itemize}
\item[1)] The non-observation of proton decay within the limit of a
lifetime of
$10^{32}$~years in the main channel p $\rightarrow \pi^0{\rm e}^+$,
excluding minimal
GUT unification.
 \item[2)] The LEP precision measurements, which are incompatible with
gauge-coupling unification for conventional minimal GUTs, but are in
reasonable
agreement with minimal supersymmetric GUTs, with supersymmetry broken at
the TeV
scale.
\item[3)] The top Yukawa coupling, unusually large with respect to the
one of other
quarks and leptons.
\item[4)] The possible resolution of the dark-matter problem, with some
of the neutral
supersymmetric particles as natural dark-matter candidates.
\end{itemize}
Although none of these facts is {\it per se} a compelling reason for
supersymmetry
and may find alternative explanations, it is fair to say that they can
all be
explained in the context of a supersymmetric extension of ordinary gauge
theories.
\vfill\eject
Whatever the final theory (strings?) for quantum gravity will be,
supergravity
\cite{ooiii} remains a deep and non-trivial extension of the principle
of general
covariance and Yang--Mills gauge symmetry, which played such an
important role in the
description of natural phenomena.

Let us hope that nature has used this fascinating structure!
\vfill\eject

 \end{document}